\documentclass{article}
\usepackage{amsmath,amsthm,amsfonts,graphicx}
\usepackage{multirow}
\usepackage{slashbox}
\usepackage{multirow}
\def\smallddots{\mathinner{\raise7pt\hbox{.}\raise4pt\hbox{.}\raise1pt\hbox{.}}}
\def\smallsdots{\mathinner{\raise1pt\hbox{.}\raise4pt\hbox{.}\raise7pt\hbox{.}}}

\numberwithin{equation}{section}
\numberwithin{table}{section}

\newtheorem{example}{Example}[section]

\newtheorem{remark}{Remark}[section]

\usepackage{graphicx}
\graphicspath{%
    {converted_graphics/}% inserted by PCTeX
    {/}% inserted by PCTeX
}
\begin{document}

%\title
\centerline{{\Large \bf Matrix Multiplication, Trilinear Decompositions,}}  

\centerline{{\Large \bf APA Algorithms, and Summation}}

%\author
\medskip

\medskip

\centerline{Victor Y. Pan}
%$^{[1, 2],[a]}$ 
%\and
%\\
%$^{[1]}$ 

\medskip

\medskip

\centerline{Department of Mathematics and Computer Science}

\centerline{Lehman College of the City University of New York}

\centerline{Bronx, NY 10468 USA}

\centerline{and}
%$^{[2]}$ 
\centerline{Ph.D. Programs in Mathematics  and Computer Science}
\centerline{The Graduate Center of the City University of New York}
\centerline{New York, NY 10036 USA}
%$^{[a]}$ 
\centerline{victor.pan@lehman.cuny.edu}
\centerline{http://comet.lehman.cuny.edu/vpan/}  
%$^{[b]}$ gqian@gc.cuny.edu \\
%$^{[c]}$ azheng-1999@yahoo.com \\
%} 

 \date{}

%\maketitle

% - - - - - - - - - - - - - - - - - - - - - - - - - - - - - - - - - - - - -

%\section{Introduction}\label{sintro}

\bigskip

% - - - - - - - - - - - - - - - - - - - - - - - - - - - - - - - - - - - - -

\begin{abstract}
Matrix multiplication (hereafter we use the acronym MM) is among the most 
fundamental operations of modern computations. The efficiency of its 
performance depends on various factors, in particular vectorization, data 
movement and arithmetic complexity of the computations, but here we focus 
just on the study of the arithmetic cost and the impact of this study on 
other areas of modern computing. In the early 1970s it was expected that 
the straightforward cubic time algorithm for MM will soon be accelerated 
to enable MM in nearly quadratic arithmetic time, with some far fetched 
implications. While pursuing this goal, the mainstream research had its focus 
on the decrease of the classical exponent 3 of the complexity of MM towards 
its lower bound 2, disregarding the growth of the input size required in order to 
support this decrease. Eventually, surprising combinations of novel ideas and
sophisticated techniques enabled the decrease of the exponent to its 
benchmark value of about 2.38, but the supporting MM algorithms  
improved the straightforward one only for the inputs of immense sizes. 
Meanwhile, the communication complexity, rather 
than the arithmetic complexity, has become the bottleneck of computations in 
linear algebra. This development may seem to undermine the value of the past 
and future research aimed at the decrease of the arithmetic cost of MM, but 
we feel that the study should be reassessed rather than closed and forgotten. 
We review 
the old and new work in this area in the present day context, recall some 
major techniques introduced in the study of MM, discuss their impact on the 
modern theory and practice of computations for MM and beyond MM, and link 
one of these techniques to some simple algorithms for inner product 
and summation. 
\end{abstract}
 

 
% - - - - - - - - - - - - - - - - - - - - - - - - - - - - - - - - - - - - -

\section{Introduction}\label{sintr}

% - - - - - - - - - - - - - - - - - - - - - - - - - - - - - - - - - - - - -

%and place the old and new study of MM into the context
% of modern computing.
 Matrix multiplication 
(hereafter we keep using the acronym {\em MM}) is fundamentally important 
for computations in linear algebra and for
the theory of computing.    
Efficient performance of MM
% may seem to be quite straightforward, but actually 
depends on various factors, 
particularly on vectorization, data locality, and arithmetic cost
(cf.\ \cite[Chapter 1]{GL13}).
We review just the work on decreasing 
the arithmetic cost
and comment on the impact of this work on modern computing beyond MM.
Our review is a revision of \cite{P84} 
 in the light of the information from the last 30 years.

The cubic arithmetic time $2n^3-n^2$ of 
the straightforward algorithm for $MM(n)$, that is,
for $n\times n$ MM,
was commonly believed to be optimal until 1969, when it
was decreased to $O(n^{2.8074})$ in \cite{S69},
implying the decrease of the classical exponent 3  
 to $2.8074$ also for numerous other venerated computational problems
such as Boolean MM and the solution of a nonsingular 
linear system of  $n$ equations. The worldwide interest to MM has
immediately exploded,
and it was widely expected that  
new efficient algorithms would soon perform MM and solve 
the related computational problems in nearly
quadratic time. Even the exponent 2.8074, however, defied
the attacks of all experts worldwide for almost 10 years, until 1978, 
and since then the mainstream research has been
directed to the decrease of
 the exponents of MM allowing 
 unrestricted input sizes. 

New surprising resources have been found, 
  sophisticated 
techniques have been developed,
and the exponents have eventually decreased  to
   the current record level of about 2.38. 
 The  overhead constants
hidden in the ``O" notation of $O(n^{2.38})$
have been notoriously huge, however, and
the supporting algorithms 
 beat the straightforward one only for
$n\times n$ input
matrices of immense sizes.
For $n$  restricted to be ``moderate",
 say, less than 1,000,000,  the current record  
is just  2.7734 of \cite{P82}, unbeaten since 1982,
%The resulting fast MM was actually only asymptotically fast,
%providing no acceleration of MM of realistic sizes.
and the prospects for serious practical impact 
are even more bleak now.
Here is a relevant citation from  \cite{BCD14}:
``The traditional metric for the efficiency of a numerical algorithm has been the number of arithmetic operations it performs. Technological trends have long been reducing the time to perform an arithmetic operation, so it is no longer the bottleneck in many algorithms; rather, communication, or moving data, is the bottleneck".

In the context of this development,
 we refocus our presentation versus  
 \cite{P84}. We realize that 
many scientists are still curious whether MM and 
numerous related computational problems can be performed 
in nearly quadratic time, even for unrestricted input size.
We cover briefly the effort motivated by this challenge
and in Section \ref{sfrth} summarize the progress.
Instead of various amazing
sophisticated techniques proposed exclusively for the
 acceleration of MM of immense sizes,
however,
%We discuss some interesting
%by-products of the research on fast MM, 
%and hope that our paper will 
%help direct future study
%of this and related subjects.
we  cover
the 
techniques
that
are efficient already for MM of moderate sizes 
or had interesting impacts
on realistic computations beyond MM. 
%that is, the main concepts and  techniques 
%introduced and advanced in this study but 
The impacts have been limited to a few items, but  
played rather important role in the past
and still have unexplored links to some apparently unrelated 
areas of computation, as
in the case of our extension of an old MM technique
  to the computation of an inner product and
summation (see Examples \ref{exinner} and \ref{exsum} in Section \ref{sbsg}).

We organize our paper as follows. In Sections \ref{s69,78}
and \ref{strex} we recall the two initial 
accelerations of MM, 
in 1969 and 1978, respectively, and comment on 
their impacts beyond MM.  
In Sections
\ref{sblnr}, \ref{strld}, and \ref{sapa} 
we cover the fundamental classes of 
 bilinear, trilinear and the so called APA algorithms, respectively,
and discuss their role in 
theoretical and practical acceleration of MM. 
In Section \ref{sfrth} we summarize the history of 
 the subject
and 
comment on its current state and prospects of its study. 
In Section \ref{sbsg} we extend a 
fundamental MM technique to  
computing inner products and summation.
%The Appendix is our translation of the paper \cite{P72} into English.
  
% - - - - - - - - - - - - - - - - - - - - - - - - - - - - - - - - - - - - -

\section{The Breakthrough in 1969: from the Exponent 3 to 2.8074}\label{s69,78}

% - - - - - - - - - - - - - - - - - - - - - - - - - - - - - - - - - - - - -

 In addition to the acronym $``MM"$ for ``matrix multiplication", 
we write $``MI"$  for ``nonsingular matrix inversion", 
 $MM(m,k,n)$  for 
the computation of the product $C=AB$ of two matrices $A$ of the size $m\times k$
and $B$  of the size $k\times n$,
$MM(n)$  for $M(n,n,n)$, and
 $MI(n)$  for $n\times n$ $MI$.
 $A=(a_{i,j})_{i,j=1}^{m,n}$
 denotes an $m\times n$ matrix with the entries 
$a_{i,j}$, for $i=1,\dots, m$ and $j=1,\dots, n$.

At first we recall the algorithm
of \cite{S69}, which 
performs $MM(n)$ for any $n$ by using $cn^{\omega}$
arithmetic operations overall, for  $\omega=\log_2 (7)<2.8074$
and a constant $c$ independent of $n$,
versus  
 $2n^3-n^2$ in the straightforward algorithm.
The basic step of  \cite{S69} is the computation of
 the product $C=AB$ of a pair of $2\times 2$ matrices, 
$$A=\begin{pmatrix} 
a_{11} & a_{12} \\ 
a_{21} & a_{22}
 \end{pmatrix},~B=\begin{pmatrix} 
b_{11} & b_{12} \\ 
b_{21} & b_{22}
 \end{pmatrix},~C=AB=\begin{pmatrix} 
c_{11} & c_{12} \\ 
c_{21} & c_{22}
 \end{pmatrix}$$ by using the following expressions, 
$p_1=(a_{11}+a_{22})(b_{11}+b_{22}),~p_2=(a_{21}+a_{22})b_{11},~p_3=a_{11}(b_{12}-b_{22})$,  \\
$p_4=(a_{21}-a_{11})(b_{11}+b_{12}),~p_5=(a_{11}+a_{12})b_{22},~p_6=a_{22}(b_{21}-b_{11})$,  
$p_7=(a_{12}-a_{22})(b_{21}+b_{22})$,  \\
$c_{11}=p_1+p_6+p_7-p_5,~c_{12}=p_3+p_5,~c_{21}=p_2+p_6,~c_{22}=p_1+p_3+p_4-p_2$.

\medskip

 The algorithm  performs 7 scalar multiplications, instead of 8 of
the straightforward algorithm, and this enables us 
to decrease the classical MM exponent 3 to 2.8074.
Indeed, assume that the entries of the matrices $A$, $B$, and $C$ above
are the $2^{p-1}\times 2^{p-1}$ blocks of a $2\times 2$ block matrix and apply
the algorithm to this block matrix. Then the scalar multiplications turn into 
multiplications of the pairs of $2^{p-1}\times 2^{p-1}$ matrices.
View them as the pairs of $2\times 2$ block matrices with 
$2^{p-1}\times 2^{p-1}$ entries and apply the same algorithm again.
Recursively apply the algorithm  
 to  the $2\times 2$ block matrices with the blocks
 of the sizes $2^{p-i} \times 2^{p-i}$, for $i=1,2,\dots,p$.
Overall $MM(2^p)$ involves only $7^p$ scalar 
multiplications rather than the straightforward $2^{3p}$. 
By embedding the $n\times n$ input matrices into 
$2^p\times 2^p$ matrices banded with zeros,
 perform $MM(n)$ by using $7^p$ multiplications in the case of any $n$
such that  $n\le 2^p<2n$.
Additions and subtractions of pairs of $n\times n$ matrices
are relatively fast, each using only $n^2$ arithmetic 
operations, and we can  
 extend the algorithm  
to performing $MM(n)$ for any $n$
at the claimed arithmetic cost $cn^{\log_2 (7)}$.
 With some additional care, one 
can yield the constant $c\approx 4.54$
and  extend
the 
 complexity bounds  
$O(n^{\log_2 (7)})$
 to
 $MI(n)$, 
the solution of a nonsingular linear system of $n$ equations,
Boolean $MM(n)$,
% more comments in the next section). 
and a number of other  
computational problems well known in 
Computer Science, Linear Algebra,  and Computer  Algebra
 (see  \cite{S69}, \cite{BH74}, 
% \cite[parts 1--3 of Theorem 1]{P72},
 \cite[Sections 6.3--6.6]{AHU74}, \cite[pages 49--51]{BM75},
and \cite[Chapter 2]{BP94}, \cite{DI00},  \cite{Z02},
 \cite{YZ04}, \cite{YZ05},  \cite{KSV06},
 \cite{AP09}, \cite{SM10},  \cite{ASU13}).

% - - - - - - - - - - - - - - - - - - - - - - - - - - - - - - - - - - - - -

\section{Bilinear Algorithms, Their Ranks and the $MM$ Exponents}\label{sblnr}

% - - - - - - - - - - - - - - - - - - - - - - - - - - - - - - - - - - - - -

The above algorithm for $MM(2)$ belongs to the important 
class of bilinear algorithms. 
Such an algorithm for $MM(m,k,n)$
 computes at first some linear forms $l_q(A)$ and $l'_q(B)$ in the entries of 
the input matrices $A=(a_{ij})_{i,j=1}^{m,k}$ and $B=(b_{jh})_{j,h=1}^{k,n}$
and then  
the entries $c_{ih}=\sum_j a_{ij} b_{jh}$ of the product $C=AB$ as 
the $mn$ bilinear forms,
$$l_q(A)=\sum_{i,j=1}^{m,k}u_{ij}^{(q)}a_{ij},~l'_q(B)=\sum_{j,h=1}^{k,n}v_{jh}^{(q)}b_{jh},~q=1,\dots,r,$$
$$c_{ih}=\sum_{q=1}^rw_{ih}^{(q)}l_q(A)l'_q(B),~~i=1,\dots,m;~h=1,\dots,n.$$
Here $r$ is said to be the {\em rank of the algorithm}, 
%its minimum over all
%bilinear algorithms for $M(m,k,n)$ is said to be the {\em rank of the $M(m,k,n)$ problem},
$u_{ij}^{(q)}$, $v_{jh}^{(q)}$, and $w_{ih}^{(q)}$ are constants, for all $i,j,h$, and $q$,
and the entries of the matrices $A$ and $B$ are variables
or  block matrices. Suppose $m=k=n$, assume that the entries are
block matrices, and
 proceed as in the previous section, 
that is, apply
recursively this bilinear algorithm of rank $r$
to block matrices. The resulting 
 bilinear  algorithms have rank $r^{p}=cn^{\omega}$ 
for
$MM(n^p)$,
$p=2,3,\dots$, and one can
extend them to perform
$MM(K)$ for any $K$ by using $cK^{\omega}$ arithmetic operations for 
$\omega=\log_{n}(r)$ and a constant $c$ independent of $K$.

The straightforward  algorithm for $MM(m,k,n)$ is bilinear of rank $mkn$. 
The algorithm of \cite{S69} for $MM(2)$, reproduced in our Section \ref{s69,78},
is bilinear of rank 7. A bilinear algorithm of rank 6 for  $MM(2)$
would have implied the exponent $\log_2(6)< 2.585$, 
but rank 7  turned out to be 
the sharp lower bound  for $MM(2)$ (see \cite{HK69}, \cite{HK71},  \cite{BD78}).

 \cite[Theorem 3]{P72} as well as \cite{dG78}
provide a formula for all bilinear algorithms of rank 7 for $MM(2)$. One of them, by Winograd (cf.\  \cite{F74}, \cite[pages 45--46]{BM75},
 \cite[Exercise 6.5]{AHU74}, or \cite{DHSS94}) performs
$MM(2)$ by using 7 scalar multiplications and
 only 15 scalar additions and subtractions, instead of 18 in 
%the algorithm of
 \cite{S69}, and this is optimal for $MM(2)$ algorithms that use 7 scalar multiplications
\cite{P76}, \cite{B95}. 
The
Winograd algorithm still implies the same MM exponent $\log_2 (7)<2.8074$, 
but its careful implementation in \cite{F74} enables the
 decrease of the overhead constant from 4.54 to 3.95.

One can define bilinear algorithms for any {\em bilinear problem}, that is, for
the computation of any set of bilinear forms, 
e.g.,  the product of two complex numbers 
$(a_1+{\bf i} a_2)(b_1+{\bf i} b_2)=(a_1b_1-a_2b_2)+{\bf i}(a_1b_2+a_2b_1)$, 
${\bf i}=\sqrt{-1}$.
The straightforward bilinear algorithm has rank 4, and here is a rank-3 bilinear algorithm,
$l_1l_1'=a_1b_1$, $l_2l_2'=a_2b_2$, $l_3l'_3=(a_1+a_2)(b_1+b_2)$,   
$a_1b_1-a_2b_2=l_1l_1'-l_2l'_2$, $a_1b_2+a_2b_1=l_3l_3'-l_1l_1'-l_2l_2'$. 
See \cite{W70}, \cite{F72}, \cite{F72a},
\cite{P72}, \cite{BD73}, \cite{HM73}, \cite{S73}, \cite{P74}, \cite{BD76}, \cite{BD78}, 
on the early study of bilinear algorithms
and see a concise exposition in \cite{BM75}. The book
\cite{W80} shows  efficient
 bilinear algorithms for the latter task and for the
highly important 
 computation  of the  product of 2
polynomials (that is, the convolution of their coefficient  vectors),
with further applications to the design of FIR-filters. 

The minimal rank of all bilinear algorithms
for a fixed bilinear 
problem such as $MM(m,k,n)$
is called the {\em rank of the problem}.
It can be bounded in terms of the minimal arithmetic cost
of the solution of the problem
 and vice versa
\cite[Theorem 1]{P72}, \cite{BM75}, but not all interesting 
algorithms for bilinear 
problems are bilinear.
The algorithm of \cite{W68} performs $MM(n)$ for any even  $n$ by
using $(0.5n+1)n^2$ scalar multiplications and
$(1.5n^2+2n-2)n$ additions and subtractions
and has been refined slightly in \cite{W70a}.
These algorithms save about 50\% of $n^3$ 
scalar multiplications, but they are not bilinear.
They use commutativity of the products of the entries of the input matrices and 
thus cannot be applied to block matrices, cannot be used recursively,
and
have made no impact on the exponent of MM.
The algorithms, however, 
 are of interest for MM of small sizes \cite{DIS11}.
(See \cite{W71} on some lower bounds on
the rank of such algorithms for MM.)

% - - - - - - - - - - - - - - - - - - - - - - - - - - - - - - - - - - - - -

\section{Bilinear Dead End  and Trilinear Exit}\label{strex}

% - - - - - - - - - - - - - - - - - - - - - - - - - - - - - - - - - - - - -

The paper \cite{S69} 
 has prompted the experts around the globe to compete
for the next breakthrough
towards the goal of performing MM in quadratic 
arithmetic time, but 
even the exponent 2.8074 
defied
all attacks 
for almost a decade, from 1969 to 1978. 
The research
in these years was at first directed
towards  devising bilinear algorithms of rank 6 for $MM(2)$ 
 and of rank 21 for
$MM(3)$.
Such algorithms would imply the decrease of the exponent 2.8074,
but rank 7  turned out to be 
the sharp lower bound  for $MM(2)$ (see \cite{HK69}, \cite{HK71},  \cite{BD78}),
 while we still do not know
whether the rank of  $MM(3)$ exceeds 20 or not.
We refer to \cite{HK69}, \cite{HK71},  \cite[Theorem 1]{P72}, and
\cite{BD73},   \cite{BD78},   
for the early lower  bounds
on the arithmetic complexity and on the rank
of the problem $MM(m,k,n)$ for any triple of $m$, $k$, and $n$ 
as well as on these bounds 
 in the case of the specific problems 
%some bounds (mostly lower bounds)
 $MM(2,2,n)$, $MM(2,3,3)$, $MM(2,3,3)$,  $MM(2,3,4)$, and
$MM(2,4,4)$. The paper  \cite{MR14} presents
the current record lower bounds on the 
rank of $MM(n)$ for all $n$ (see \cite{B89},  \cite{B99},  \cite{B00},
 \cite{RS03}, \cite{S03}, \cite{L14},
on some preceding works), while the papers
\cite{DIS11} and \cite{S13} cover various lower and upper bounds
on the arithmetic complexity and the rank of MM of smaller sizes.

The stalemate in 1969--1978
ended when the paper \cite{P78} presented 
a bilinear algorithm of rank 143,640 for $MM(70)$. This implied 
the exponent $\omega=\log_{70} (143,640)<2.7962$
for $MM(n)$, $MI(n)$, Boolean $MM(n)$, 
and various other computational problems.
The algorithm of \cite{P78} has
extended an algorithm of 
the paper \cite{P72} of 1972,  
published in Russian\footnote {Until 1976 the author lived in the Soviet Union. 
From 1964 to 1976 he has been working
 in Economics 
to make his living
and has written the papers \cite{P66} and \cite{P72} in his spare time.}
%Since 1977 he has been working in the USA.}
and translated into English only in 2014
in \cite{P14}. 
 At first the paper 
 reduced the acceleration of MM  
to the decomposition of a certain trilinear form  
into the sum of fewer trilinear products,
which was a simple but novel step.
Then the paper introduced a
 nontrivial technique of 
{\em trilinear aggregation} for generating 
such decompositions. 

The technique has been the basis of the algorithm of 
\cite{P78}  and has become indispensable 
for almost all subsequent decreases of the MM exponent,
but the paper \cite{P72} was also a historical  
landmark in the study of multilinear and tensor decompositions.
Such decompositions have been introduced by Hitchcock
in 1927, but received scant attention until
a half of a dozen of papers appeared in 1963--70 
in the psychometrics literature. 
The paper \cite{P72} of 1972 was the next significant step:
 it presented the earliest known application of nontrivial 
multilinear and tensor decompositions to fundamental matrix computations.
The paper has rarely  been cited in the Western literature on MM
and never in the works on multilinear and tensor decompositions,
even though
%including 3 novelties (besides 
%its cited lower bounds on MM), that is, it (i) mapped bilinear MM problems to 
%trilinear decompositions, (ii) exploited this map to 
%reduce rectangular MM to square MM, and 
%(iii) proposed 
%the novel nontrivial technique of trilinear aggregation,
%which enabled trilinear decompositions for MM
%by using fewer terms and consequently enabled new fast MM.  
by now such decompositions have become a popular tool  
in linear and multilinear algebra and have a wide range of 
important applications to modern computing 
(see  \cite{T03}, \cite{KB09},  \cite{OT10}, 
\cite{GL13},
and the bibliography therein). 

% - - - - - - - - - - - - - - - - - - - - - - - - - - - - - - - - - - - - -

\section{Trilinear Decompositions and Trilinear Aggregation}\label{strld}

% - - - - - - - - - - - - - - - - - - - - - - - - - - - - - - - - - - - - -

As we said, the paper \cite{P72} proposed  and  exploited
trilinear representation in order to accelerate MM. 
Next we outline this approach and link it to some important
computations beyond MM.

At first assume a bilinear algorithm  of rank $r$ 
for a matrix product $C=AB$ of an $m\times k$
matrix $A$ by a $k\times n$
matrix $B$. Multiply  its equations
$\sum_j a_{i,j} b_{jh}=\sum_{s=1}^rw_{ih}^{(s)}l_s(A)l'_s(B)$
by new variables $d_{hi}$, sum the products in $i$ and $h$,
and  arrive at 
a  {\em trilinear decomposition} of rank $r$
for $trace(ABD)=\sum_{i,j,h} a_{i,j} b_{jh}d_{hi}$ where $D=(d_{hi})_{h,i}^{n,m}$
is an auxiliary  $n\times m$ matrix and 
 $trace(M)$ denotes the trace of a matrix $M$.
(This is equivalent to the decomposition of rank $r$ for the tensor 
 associated to the trilinear form $trace(ABD)$, that is, to its 
 decomposition
into the sum of $r$ tensors of  rank 1.)
For example, 
here are the trilinear extensions of the bilinear algorithms
of the previous section, that is, a trilinear
decomposition 
of rank 3 for 
 multiplication
of two complex numbers, 
$$a_1b_1d_1-a_2b_2d_1+a_1b_2d_2+a_2b_1d_2=
a_1b_1(d_1-d_2)-a_2b_2(d_1+d_2)+(a_1+a_2)(b_1+b_2)d_2,$$
and a trilinear decomposition    
of rank 7 for $MM(2)$,  \\
$~~~~~~~~~ \sum_{i,j,h=1}^2a_{ij}b_{jh}d_{hi}=\sum_{s=1}^7l_sl_s'l_s''$,~
$l_1l_1'l_1''=(a_{11}+a_{22})(b_{11}+b_{22})(d_{11}+d_{22})$, \\
$l_2l_2'l_2''=(a_{21}+a_{22})b_{11}(d_{21}-d_{22})$,
$l_3l_3'l_3''=a_{11}(b_{12}-b_{22})(d_{12}+d_{22})$,
$l_4l_4'l_4''=(a_{21}-a_{11})(b_{11}+b_{12})d_{22}$,\\
$l_5l_5'l_5''=(a_{11}+a_{12})b_{22}(d_{12}-d_{11})$,
$l_6l_6'l_6''=a_{22}(b_{21}-b_{11})(d_{11}+d_{21})$,
$l_7l_7'l_7''=(a_{12}-a_{22})(b_{21}+b_{22})d_{11}$.

\medskip

Conversely, interpret both sides of a decomposition 
of the  trilinear form $trace(ABD)$
as linear forms in the variables $d_{ih}$,
equate the coefficients of these variables on both sides of the 
decomposition, and come back to the original bilinear algorithm
for  the matrix product $C=AB$. 
By formalizing the above observations, \cite[Theorem 2]{P72} 
states the equivalence of the design of
a bilinear algorithm of rank $r$ for $MM(m,k,n)$
 to a trilinear decomposition 
of rank $r$ for 
 the associated trilinear form and its tensor.

By equating the coefficients of 
all variables $a_{ij}$ or all variables $b_{jh}$ on both sides 
of the trilinear decomposition, we obtain 2 
other dual bilinear algorithms
of the same rank
for the problems $M(k,n,m)$ and $M(n,m,k)$. 
By interchanging the subscripts of the variables,
we arrive at the dual bilinear algorithms of the same rank  
for the problems $MM(m,n,k)$, $MM(k,m,n)$, and $MM(n,k,m)$ as well
(cf.\ \cite[part 5 of Theorem 1]{P72},  \cite{BD73},
\cite{HM73}, \cite{P74}). 
The latter extension from triples to 6-tuples 
of algorithms uses the double subscripts 
for matrix entries and is peculiar to MM, 
but the triples of bilinear algorithms
can be generated from their common 
trilinear representation 
for any bilinear computational problems,
and
the book \cite{W80}
 applies this technique in order
 to devise new  efficient bilinear algorithms
for FIR-filters and multiplication of complex numbers and polynomials.
 
For another 
demonstration of this duality technique, let us deduce 
 part 1 of \cite[Theorem 1]{P72}, which states 
that we can perform $MM(K)$ by using $cK^{\omega}$
arithmetic operations for
 the exponent $\omega=3\log_{mkn}(r)$
and a constant $c$ independent of $K$
provided that we are given a bilinear or trilinear algorithm 
of rank $r$ for  rectangular $MM(m,k,n)$
for any specific dimensions $m$, $k$, and $n$.
In Section \ref{sblnr} we have already deduced this result in the case where $m=k=n$.
Now successively apply the 3  dual bilinear algorithms of rank $r$
  to block 
 $MM(m,k,n)$, block $MM(n,m,k)$,
and  $MM(k,n,m)$ and arrive at
 a bilinear algorithm of rank $r^3$ for
$MM((mkn)^3)$.
Then the claim follows from the result of Section \ref{sblnr}
for $n$ replaced by $(mkn)^3$. 

Motivated by the basic fact of its Theorem 2
about the  equivalence of 
bilinear and
trilinear 
decompositions,
the paper \cite{P72} has
 introduced by an example
a nontrivial technique of 
trilinear aggregation for 
devising trilinear decompositions of small rank
for  
$trace(ABD)$. 
Here is a similar  example.
At first  
assume that we seek
two independent matrix products $AB$ and $UV$
and associate the pair of disjoint trilinear forms,
$trace(ABD+UVW)=\sum_{i,j,h=1}^{m,k,n}(a_{ij}b_{jh}d_{hi}+u_{jh}v_{hi}w_{ij})$
to  this task of $Disjoint~MM$.
Sum the trilinear aggregates 
$S=\sum_{i,j,h=1}^{m,k,n}(a_{ij}+u_{jh})(b_{jh}+v_{hi})(d_{hi}+w_{ij})$
and then subtract the correction terms $T_1=\sum_{i,j=1}^{m,k}a_{ij}q_{ij}w_{ij}$,
$T_2=\sum_{j,h=1}^{k,n}u_{jh}b_{jh}r_{jh}$, and $T_3=\sum_{h,i=1}^{n,m}p_{ih}v_{hi}d_{hi}$,
where $p_{ih}=\sum_{j=1}^{k}(a_{ij}+u_{jh})$,
$q_{ij}=\sum_{h=1}^{n}(b_{jh}+v_{hi})$, and $r_{jh}=\sum_{i=1}^{m}(d_{hi}+w_{ij})$.
The resulting equation $trace(ABD+UVW)=S-T_1-T_2-T_3$ defines
a trilinear decomposition
 of rank $mkn+mk+kn+nm$ (rather than the straightforward $2mkn$).
Every  aggregate is the products of 3 distinct multiplicands, each 
 being the sum  of
2 entries of the same  column  in the first Generating Table below.
The correction terms are made up of the cross-products 
of the triples of the entries
from 3 distinct columns of the table
such that the 3 entries do not lie in the same row.

The technique of 
trilinear aggregation is naturally linked to Disjoint MM
(see \cite{P84},  \cite[Section 5]{P84a},
 \cite[Section 12]{P84b}, and \cite{LPS92} on this link),
and one can readily devise
fast Disjoint MM of reasonable sizes by using
this link.
The above construction of Disjoint MM, a similar one below, 
and the ones of \cite{LPS92}, however, have been quite 
 readily extended to $MM(n)$. 
In particular, by playing with odd and even subscripts
of the matrix entries,
the paper \cite{P72}
has 
extended 
the above decomposition of $trace(ABD+UVW)$ to
 a bilinear algorithm of rank $0.5 n^3+3n^2$ for $MM(n)$ and any even $n$.
This implied the $MM$ exponent $\log_n (0.5n^3+3n^2)$,
which is less than 2.85 for  $n=34$. 
 
By using 
 the second and the third Generating Tables below,
we can decrease this exponent further.
Indeed sum the $mkn$
aggregates 
$(a_{ij}+u_{jh}+x_{hi})(b_{jh}+v_{hi}+y_{ij})(d_{hi}+w_{ij}+z_{jh})$
of the second Generating Table,
subtract order of $n^2$ correction terms,  and
obtain a decomposition of rank $n^3+O(n^2)$
for $trace(ABC+UVW+XYZ)$,
versus the straightforward $3n^3$.
The trace represents 
3 disjoint problems of $MM(n)$, that is,
the computation of
the 3 independent matrix products $AB$, $UV$,  and $XY$,
and we obtain a bilinear algorithm of rank $n^3+O(n^2)$
for this bilinear task.
Elaboration upon this idea in \cite{P78} has resulted in
a trilinear decomposition and bilinear algorithms
 of rank $(n^3-4n)/3+6n^2$ 
for  $MM(n)$, $n=2p$, and any positive integer $p$.
Substitute $n=70$ and obtain the MM exponent  2.7962,
 cited in the previous section.
 
\medskip

\begin{tabular}{ c }
% $a_{ij}$ & $b_{jh}$ & $\lambda^2 d_{hi}$ 
{GENERATING} \\
% $\lambda u_{jh}$ &  $\lambda v_{hi}$   & $w_{ij}$ 
TABLES:\\ 
% $x_{hi}$   & $y_{ij}$ & $z_{jh}$ \\ \hline 
% $256$ & $1$ & $4.8\times 10^{-15}$ & $9.2\times 10^{-10}$ & $8.6\times 10^{-11}$ & $2.1\times 10^{-11}$ \\ \hline
    \end{tabular}  ~~
    \begin{tabular}{| c | c |c |}
      \hline
 $a_{ij}$ & $b_{jh}$ & $d_{hi}$  \\ \hline
 $u_{jh}$ &  $v_{hi}$   & $w_{ij}$ \\ \hline
% $x_{hi}$   & $y_{ij}$ & $z_{jh}$ \\ \hline 
% $256$ & $1$ & $4.8\times 10^{-15}$ & $9.2\times 10^{-10}$ & $8.6\times 10^{-11}$ & $2.1\times 10^{-11}$ \\ \hline
    \end{tabular}
~~
%------------------------------------------------------------------------------
%\begin{table}[ht]
  %\caption{2-term and 3-term aggregates}
 % \label{tab2agg}
 % \begin{center}
    \begin{tabular}{| c | c |c |}
      \hline
 $a_{ij}$ & $b_{jh}$ & $d_{hi}$  \\ \hline
 $u_{jh}$ &  $v_{hi}$   & $w_{ij}$ \\ \hline
 $x_{hi}$   & $y_{ij}$ & $z_{jh}$ \\ \hline 
% $256$ & $1$ & $4.8\times 10^{-15}$ & $9.2\times 10^{-10}$ & $8.6\times 10^{-11}$ & $2.1\times 10^{-11}$ \\ \hline
    \end{tabular}
 % \end{center}
~~~
\begin{tabular}{| c | c |c |}
      \hline
 $a_{ij}$ & $b_{jh}$ & $\lambda^2 d_{hi}$  \\ \hline
 $\lambda u_{jh}$ &  $\lambda v_{hi}$   & $w_{ij}$ \\ \hline
% $x_{hi}$   & $y_{ij}$ & $z_{jh}$ \\ \hline 
% $256$ & $1$ & $4.8\times 10^{-15}$ & $9.2\times 10^{-10}$ & $8.6\times 10^{-11}$ & $2.1\times 10^{-11}$ \\ \hline
    \end{tabular}
%\end{table}

\medskip

%This is above 2.8074 of \cite{S69},
%but below the classical $MM$ exponent 3.
%the earliest  known application
%of trilinear (that is, tensor)
% decompositions to 
%fundamental 
%matrix computations. 
%Now application of tensor decompositions to
%matrix computations is a thriving area of scientific computing 
%(see [KB09], [OT10], and the bibliography therein).  

% - - - - - - - - - - - - - - - - - - - - - - - - - - - - - - - - - - - - -

\section {APA Algorithms and Further Acceleration of $MM$}\label{sapa}

% - - - - - - - - - - - - - - - - - - - - - - - - - - - - - - - - - - - - -
 
The third Generating Table helps us to demonstrate the
technique of
{\em Any Precision Approximation} 
 introduced in  \cite{BCLR79} and \cite{B80}
(hereafter we use the acronym 
{\em APA}).
Its combination with trilinear aggregation and 
Disjoint MM has
enabled further 
%substantial 
decrease of 
  the MM exponents. Moreover,
our Section \ref{sbsg}
and  \cite[Section 40]{P84b} link the APA technique
to some fundamental computations beyond MM.
%APA algorithms become a basic 
%part in the study of tensor decompositions.

For a sample APA algorithm, generate the  sum 
of $mkn$ aggregates
$S=\lambda^{-1} \sum_{i,j,h=1}^{m,k,n}(a_{ij}+\lambda u_{jh})(b_{jh}+
\lambda v_{hi})(\lambda^2 d_{hi}+w_{ij})$
 from the third Generating Table above. 
For $\lambda=1$ this table turns into the first Generating Table,
and then trilinear aggregation of the previous section 
enables a decomposition of rank $mkn+mk+kn+mn$ for $trace(ABD+UVW)$.
Suppose, however, that $\lambda\rightarrow 0$ and 
obtain a trilinear decomposition $trace(ABD+UVW)=S-T_1-T_2+O(\lambda)$,
where
 $T_1=\sum_{i,j=1}^{m,k}a_{ij}q_{ij}w_{ij}$ and
$T_2=\sum_{j,h=1}^{k,n}u_{jh}b_{jh}r_{jh}$,
for
$q_{ij}=\sum_{h=1}^{n}(b_{jh}+\lambda v_{hi})$
 and $r_{jh}=\sum_{i=1}^{m}(\lambda^2 d_{hi}+w_{ij})$.
If we ignore the terms of order $\lambda$,
then the rank of the decomposition turns into its 
$\lambda$-{\em rank} 
 or {\em  border rank} 
%by Bini and Capovani) 
and decreases to $mkn+mk+kn$. 

By equating the coefficients of the variables $d_{hi}$
and $w_{ij}$ in the trilinear APA decomposition above,
we arrive at the bilinear problem of the evaluation of 
two disjoint matrix products $AB$ and $UV$. The $mkn$  
trilinear aggregates turn into the $mkn$  
 bilinear  products
 $(a_{ij}+\lambda u_{jh})(b_{jh}+
\lambda v_{hi})$ for all $i$, $j$, and $h$.
Clearly, $a_{ij}+\lambda u_{jh}\rightarrow a_{ij}$
and $b_{jh}+
\lambda v_{hi} \rightarrow b_{jh}$
as $\lambda \rightarrow 0$, 
but we must keep the terms   
$\lambda u_{jh}$ and $\lambda v_{hi}$
in the aggregates
in order to compute the matrix  product $UV$.
This can force us to double the precision of the  
representation of
the multiplicands   
 $a_{ij}+\lambda u_{jh}$
and $b_{jh}+
\lambda v_{hi}$
compared to the precision of the representation
of the entries 
 $a_{ij}$, $u_{jh}$, $b_{jh}$, and
$v_{hi}$.
E.g., suppose that $\lambda=2^{-s}$
for a sufficiently large integer $s$ and that
$a_{ij}$, $u_{jh}$
 $b_{jh}$, and
$v_{hi}$ are $s$-bit integers in the range $[0,2^s)$.
Then $2s$ bits
are required to represent
each of the multiplicands   
 $a_{ij}+\lambda u_{jh}$
and $b_{jh}+
\lambda v_{hi}$.

One would prefer to avoid doubling the precision 
if $s$ exceeds  the  
half-length of the computer word,
%that is, of the allowed precision of the computations, 
but in \cite{B80} Bini proved that
we can ignore this problem
if our goal is just the decrease of the exponents of MM.
%Namely, we can recursively apply the algorithm 
%to block MM as many times as we wish,
Indeed, multiply the above trilinear decomposition of 
$trace(ABD+UVW)$ by 
the  variable $\lambda$
 and arrive at a trilinear decomposition whose
coefficients are polynomials in $\lambda$ of degree $d=2$.
In particular $trace(ABD+UVW)$ is the coefficient of $\lambda$
in this decomposition, which we can compute by means of
interpolation.
 
For  decreasing the exponent of MM,
this observation is not sufficient yet,
 because the interpolation stage increases the
 rank  by a factor of  $(d+1)^2$,
that is, by a factor of 9 for $d=2$. In the above case
%With FFT we can decrease this factor to $O(d\log(d))$.
  the resulting rank $9(mkn+mk+kn)$ 
exceeds the rank $2mkn$ of the straightforward algorithm
for $trace(ABD+UVW)$.  
After 
sufficiently many recursive applications of the algorithm,
however, the comparison  is reversed
because in every recursive step the degree of a decomposition in 
$\lambda$ is only doubled, whereas the matrix size is squared, that is, 
the degree grows logarithmically in the matrix size.
Ultimately the impact of the interpolation factor on the MM 
exponent becomes immaterial.
 Hence the  MM exponent $3\log_{mkn}r$ 
is defined even by an APA 
 trilinear decomposition of $trace(ABD)$ having border rank $r$
if $trace(ABD)$
represents $MM(m,k,n)$ for  fixed  $m$, $k$, and $n$.

The APA decomposition above for 
$trace(ABD+UVW)$ is associated with Disjoint MM rather than MM,
but Sch{\"o}nhage in \cite{S81} proved that
the exponents of MM can be obtained readily from 
the decompositions for  Disjoint MM as well.
In particular the above APA decomposition of $trace(ABD+UVW)$
implies the MM exponent $\omega=3\log_{mkn}(0.5(mkn+mk+kn))$ for any triple
of $m$, $k$, and $n$.
Substitute $m=n=7$ and $k=1$ and obtain 
the MM exponent $\omega=3\log_{49}31.5<2.66$.
Further refinement of this construction in  \cite{S81} 
yielded the exponent
$\omega<2.548$, and \cite{P81} decreased this record to $\omega<2.522$
by  combining APA algorithms
and trilinear aggregation 
 for disjoint MM
 represented by $trace(ABC+UVW+XYZ)$.

By continuing this line of research, the next 2 celebrated papers \cite{S86}
and \cite{CW90} decreased the exponents to the  record of about 2.38,
which was further decreased by 0.002 in 2010 and 0.001 in 2012--2014, respectively.
Namely, due to the  results in \cite[part 1 of Theorem 1
and Theorem 2]{P72}, \cite{B80},  \cite{S81},  \cite{S86}, and  \cite{CW90}
it became possible and challenging to deduce  
small exponents of MM  
from more and more lenient basic bilinear or trilinear decompositions 
of small rank for  MM and
 Disjoint MM of small sizes, and since 1986 for some small
bilinear 
%or even for some bilinear or trilinear
problems  similar to Disjoint MM.
All these results have been
built on the top of  the techniques of 
the preceding papers and
have been increasing in sophistication. Accordingly they  
required performing  
 longer recursive
processes of  nested block MMs, 
and the input size had to 
be blown up 
to accommodate these processes.
This deficiency has been avoided by 
 trilinear aggregation, but by none of
the subsequent techniques.

 % - - - - - - - - - - - - - - - - - - - - - - - - - - - - - - - - - - - - -

\section{A Summary of the Research on Fast MM}\label{sfrth}

% - - - - - - - - - - - - - - - - - - - - - - - - - - - - - - - - - - - - -

Let us summarize. Strassen's seminal paper \cite{S69} of 1969 has
motivated extensive effort worldwide with the goal of decreasing the exponent of MM
towards its information lower bound 2. In spite of the initial optimism, 
almost a decade of stalemate followed.
Then,  since 1978, application of trilinear aggregation 
and,  since 1979,
its combination with the  techniques of APA algorithms
and Disjoint MM  enabled  
further decrease of the MM exponents.  
Strassen's exponent 2.8074 of 1969 has been beaten
a number of times in 1978--1981, ending with 2.496
of  \cite{CW82}. Two new decreases in 1986 produced the
benchmark record  2.376 of \cite{CW90},
 decreased  by tiny margins, 
to  2.374 in \cite{S10} and \cite{DS13}
and 2.373  in \cite{VW14} and \cite{LG14}. (The title of \cite{VW14}
 is deceptive a little. The paper has decreased the fresh exponent of \cite{S10} and \cite{DS13}, but
not the decades-old exponent of  \cite{CW90}.)
Almost all decreases of the MM exponents
 relied on amazing novel techniques built on the top of each other.
According to
 \cite[page 255]{CW90}, all of them employed trilinear 
aggregation and since 1979 combined it with
APA algorithms.

% After 1981 and particularly after 1987,
%  the progress slowed down.
%Moreover, since 1979 the decrease of the exponents has lost any touch 
%to practical MM
%because all this progress was achieved only 
% for the matrices of immense sizes. 
%Trilinear aggregation and the 
%widely acclaimed
%group-theoretic approach of \cite{CU03} to  fast MM were the only 2 exceptions,
%although so far the acceleration of MM by using the approach 
% of \cite{CU03} has never reached even 
%the level of 1969 of \cite{S69}.
 
Already in his paper \cite{S81} of 1981, however,
Sch{\"o}nhage pointed out 
that all new exponents of MM were just "of theoretical interest"
because they were valid only 
for the inputs "beyond any practical size", and so, according to \cite{S81}, 
"Pan's estimates of 1978 for moderate" input sizes  
were "still unbeaten".  Actually 
the exponent 2.7962 of 1978 for  
$MM(n)$ (restricted, say, to $n\le 1,000,000$)
has successively been decreased in \cite{P79},
 \cite{P80},  \cite{P81}, and  \cite{P82}  (cf.\ also \cite{P84}), 
although by  
small margins. 
As of November 2014, the exponent 2.7734 of  \cite{P82} 
is still the record low for $MM(n)$ with $n\le 1,000,000$. 

\begin{figure}[tbp]  
%float placement: (h)ere, page (t)op, page (b)ottom, other (p)age
  \centering

  \includegraphics[width=6in,height=3in,keepaspectratio]{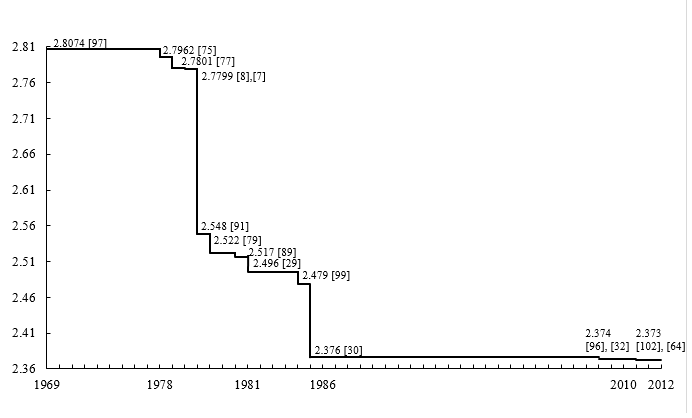}
  \caption{$MM(n)$ exponents for unrestricted $n$.}
  \label{fig1}
\end{figure}

\begin{figure}[h]
	\caption{$MM(n)$ exponents for $n\le 1,000,000$.}
	\label{fig2}
		\centering
			\includegraphics[width=6in,height=3in,keepaspectratio]{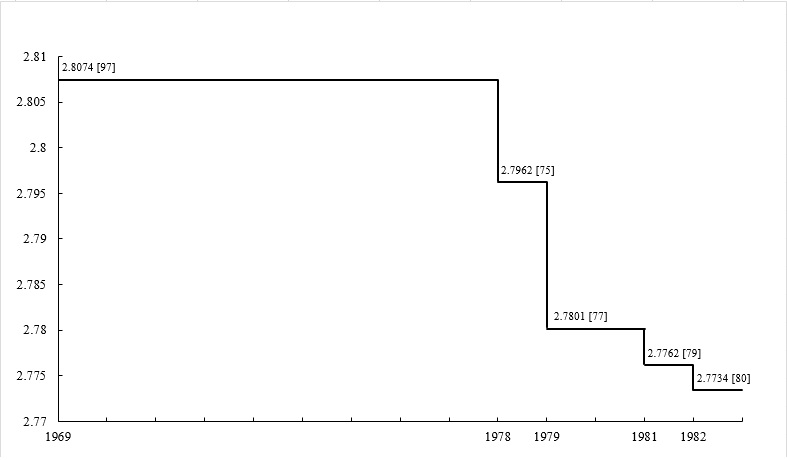}
\end{figure}

Tables \ref{tab1} -- \ref{tab2} and Figures \ref{fig1} and 
\ref{fig2} show 
 the dynamics of the exponents of $MM(n)$
 since 1969, under no restriction on the dimension $n$ 
and for $n\le 1,000,000$, respectively.  
The tables link each exponent to its recorded
publication in a journal, a conference proceedings, or
as a research report.
Accordingly, Figures \ref{fig1} and 
\ref{fig2} display the chronological process and reflect
the competition for the decrease of the MM exponents, 
particularly intensive in 1979--1981 and 1986.  
The record exponents of MM have been 
updated 4 times during the single year of 1979.
At first the MM exponent 2.7801 appeared in February in a Research Report 
(see \cite{P80}). 2.7799  
appeared as an APA MM exponent
in \cite{BCLR79} in June
and then as an MM  exponent
in \cite{B80}.
 The next
MM exponent 2.548 of \cite{S81} was followed by the MM 
exponent 2.522 of \cite{P81}, both published in the   
book of abstracts 
of the conference on the Computational Complexity in Oberwolfach, West Germany, 
organized by Schnorr, Sch{\"o}nhage and Strassen 
(cf.\ \cite[page 199]{P84} and \cite{S81}). 
The exponent 2.496 of \cite{CW82}
 was reported in October 1981 at the IEEE FOCS'81,
but in Table \ref{tab1} and Figure
\ref{fig1} we place it after the exponent 2.517
of the paper \cite{R82} of 1982,
which was submitted in March 1980.
The Research Report version of the paper \cite{CW90}
appeared in August of 1986, 
but in Table \ref{tab1} and Figure
\ref{fig1} we place  \cite{CW90} after the paper  
\cite{S86},  published in October of 1986 in the Proceedings of the IEEE FOCS,
because  the paper \cite{S86} has been  
submitted to FOCS'86 in the Spring of 1986 and has been
widely circulated afterwards.
One could complete the historical account
of Tables \ref{tab1} and \ref{tab1a} and Figure
\ref{fig1}
by including our  
exponent
 2.7804 (announced in
the Fall of 1978 but
 superseded in February 1979 in \cite{P80})
and the value 2.5218007, which
 decreased our exponent 2.5218128 of 1979  
and appeared at the end of
the final version of 
 \cite{S81}  in 1981,
that is, before the publication, but after the submission of the 
exponent 2.517 of \cite{R82}. 
Table \ref{tab1a} shows the decrease of the exponents in 1986 -- 2014.
%and  the paper \cite{CW90} cites 
%\cite{S86} as its valuable predecessor.
 
We refer the reader 
to  \cite{C82}, \cite{LR83},  \cite{C97},  \cite{HP98}, \cite{KZHP08}, 
  \cite{LG12}, and the references therein
for similar progress in asymptotic acceleration of rectangular MM
and its theoretical implications. 

\begin{table}[h] 
\caption{$MM(n)$ exponents for unrestricted $n$.}
\label{tab1}
  \begin{center}
    \begin{tabular}
{| c |c | c| c |c|c| c| c| c| c| c|c |}
      \hline
%$\omega$ & 
2.8074 & 2.7962  & 2.7801 & 2.7799 & 2.548 &2.522 & 2.517  & 2.496& 2.479 & 2.376  
& 2.374 & 2.373  \\ \hline
% paper & 
 \cite{S69}   & \cite{P78} & \cite{P80} & \cite{BCLR79}, \cite{B80} & \cite{S81}
 &\cite{P81} & \cite{R82} & \cite{CW82} &  \cite{S86} & \cite{CW90}  & \cite{S10}, \cite{DS13} & \cite{VW14}, \cite{LG14} \\ \hline
1969  &  1978  &  1979  &  1979  & 1979 & 1979 & 1980  & 1981  &    1986  &   1986  &  2010  & 2012  \\ \hline
    \end{tabular}
\end{center}
\end{table}

\begin{table}[h] 
\caption{$MM(n)$ exponents for unrestricted $n$ in 1986--2014.}
\label{tab1a}
  \begin{center}
    \begin{tabular}
{| c| c |c | c| }
      \hline
%$\omega$ & 
2.3754770 & 2.3736898  
& 2.3729269 & 2.3728639 \\ \hline
% paper 
 \cite{CW90}& 
 \cite{S10}, \cite{DS13} & \cite{VW14} & \cite{LG14} \\ \hline
1986 &  2010  & 2012\footnote{Actually, the 
proceedings version of  \cite{VW14} in STOC 2012 
claimed a bound
below 2.3727, but this turned out to be premature} & 2014 \\ \hline
    \end{tabular}
\end{center}
\end{table}

\begin{table}[h] 
\caption{$MM(n)$ exponents for $n\le 1,000,000$.}
\label{tab2}
  \begin{center}
    \begin{tabular}
{| c |c | c| c |c| }
      \hline
%$\omega$ & 
2.8074 & 2.7962  & 2.7801 & 2.7762 & 2.7734   \\ \hline
% paper & 
 \cite{S69}   & \cite{P78} & \cite{P80} & \cite{P81} &   \cite{P82}  \\ \hline
 1969   & 1978 & 1979 & 1981 &   1982  \\ \hline
    \end{tabular}
\end{center}
\end{table}

%\begin{figure}[h]
%	\caption{$MM(n)$ exponents for $n\le 1,000,000$.}
%	\label{fig2}
%		\centering
%			\includegraphics[width=1\textwidth]{F_Restricted_N}
%\end{figure}

\medskip

%In sum, the progress in decreasing the exponents of $MM(n)$ has almost stopped after 1987. 
%Moreover, in the case of dimensions $n$ not exceeding 1,000,000, the progress has stopped 
%completely  after 1982.

The decrease of the MM exponents 
displayed in Tables  \ref{tab1} and \ref{tab1a} and Figure
\ref{fig1} has been a major issue in Computer Science since 1969 and
attracted a lot of attention worldwide. Only a very 
small part of the proposed algorithms,
however, and nothing produced 
after 1971, have been used for practical MM so far. 
It should be clarified right away that, in spite of some initial fears
and persistent reservations about the early algorithms by Strassen and Winograd, 
all fast MM algorithms can be
implemented with no serious numerical stability problems
(see \cite{BL80}, \cite{DDHK07}, \cite{DDH07}).
Still the cited comments from 
Sch{\"o}nhage's paper \cite{S81} apply to all the
 acclaimed MM algorithms,
supporting the record exponents since 1979.
  
Table \ref{tab2} and Figure
\ref{fig2} display the chronology of decreasing 
the record exponents of $MM(n)$ for $n\le 1,000,000$. 
The associated overhead constants are small
because the
supporting algorithms 
avoid  recursive application of nested block MM and
rely just on the trilinear 
aggregation techniques of \cite{P72},
\cite{P78},
and \cite{P81}.
Some of these algorithms have also been refined in 
\cite{P84}, \cite{LPS92} and  then
implemented
  in \cite{K99} and \cite{K04}.
%They
% have not been used in practice, although
%according to \cite{K04},
%the implementations require less memory
%and are more stable numerically than 
%the known implementations used in practice.

The MM  algorithms used in practice (cf.\ \cite{B88}, \cite{H90},
\cite{DHSS94}, \cite{DGP04}, \cite{DGP08}, \cite{BDPZ09},
\cite{DN09}, the references therein, and
the detailed account  in \cite[Chapter 1]{GL13})
 either 
rely just on the straightforward algorithm or 
 employ
the Strassen 
and Winograd algorithms 
 supporting the MM exponent $\log_2(7)\approx 2.8074$ and
in some cases
combined with the  algorithms of \cite{W68} and \cite{W70a},
applied to the auxiliary
 matrices of smaller sizes.
These implementations make up a relatively small, but valuable part of the 
present day MM software. Kaporin's implementations 
  in \cite{K99} and \cite{K04} promise further benefits.
In particular they use small memory space, and in 
part (ii) of our comments below we point out another important 
resource for further progress. 

Even though the impact 
of the study of fast MM on computational practice 
has been limited 
and in spite of the cited decrease of the importance of the 
arithmetic complexity versus communication  complexity
of the computations in linear algebra
\cite{BCD14}, we feel that further study of 
the arithmetic complexity of MM can still be interesting if one
reassesses its goals.

(i) People  still hope to see 
 MM performed in quadratic arithmetic time up to a 
polylogarithmic factor. 
The recent 
 decrease of the old exponent from 2.376 of \cite{CW90} to about 2.373
in \cite{S10},
 \cite{DS13},  \cite{VW14}, and  \cite{LG14}
has raised some enthusiasm, but the papers \cite{ASU13} and \cite{AFLG14} have  
showed that this progress 
cannot lead  far unless it is enriched with dramatically new ideas. 

For a while 
the group theoretical approach to  MM
in  \cite{CU03} and  \cite{CKSU05}
has been regarded highly promising.
The hope was even for practical MM in nearly quadratic time, but
 the
dream has not come true, 
and  so far the approach has  produced no new
 exponents.
Moreover, the papers  \cite{ASU13} and \cite{AFLG14} have
showed some obstacles for it as well.
It is still linked to some intriguing open problems that can justify its further study, and
the paper \cite{CU13} has proposed  
 to enhance its 
 power 
by employing more general algebraic structures.

The most recent work
 \cite{AFLG14} suggested the alternative of ``finding new
identities" similar to the ones of  \cite{CW90}, that is,
coming back to the search (possibly  computer assisted) of  a
basic bilinear or trilinear decomposition 
of small rank for a task of small size that would
generalize Disjoint MM.
Then again the hope is that the extremely high tensor powers
of such a decomposition {\em can possibly
produce the desired decrease of the exponent of MM of immense sizes}.

One should, however, recall the caveat by \cite{S81} and avoid citing 
practical computing as the motivation for this advanced study. 
  
(ii) The combination of human and computer research towards
 devising efficient algorithms for MM of reasonable sizes has good chances 
for success in the future.
Such a combination has produced the seminal algorithm of \cite{BCLR79},
according to \cite{R79},
and (combined with some 
relatively simple optimization technique) has
recently enabled the acceleration of
 the known algorithms for some
rectangular MM in
the cases of
practical interest, namely for
$MM(m,k,n)$
 where $\max\{m,k,n\}\le 6$
and $\min\{m,k,n\}\le 3$  (see \cite{S13}). 
One can try to combine the optimization technique of \cite{S13}
with trilinear aggregation and computerized search in order to accelerate
 Disjoint MM  of small sizes.
The study in \cite{S13}
has not determined the rank of $M(3,3,3)$ and  has not beaten the exponent
2.7734 of  \cite{P82}, but 
 for some small rectangular MM problems has
produced algorithms that run  faster 
than the Strassen's, Winograd's, and all other known algorithms. 
This can already serve as a basis for some improvement of 
the present day MM software, but 
more advanced optimization techniques
supported by more powerful computers 
may lead to new efficient algorithms
for rectangular MM and  Disjoint MM of small sizes,
to decreasing the exponent 2.7734  
of $MM(n)$ for $n\le 1,000,000$, unbeaten since 1982,
and to revealing some powerful MM techniques. 

  (iii) The research on MM had valuable by-products for some related subject
areas. In addition to  the impact of MM on the theory of computing, which we
pointed
out  at the end of Section \ref{s69,78}, 
%let us recall that
MM has served as a basic problem in the field of 
algebraic computations, important for both theory and practice of computing (cf.\
\cite{AHU74}, \cite{BM75}, \cite{BP94}, \cite{BCS97}, \cite{K98}, \cite{GG13}). 
The origin of the field can be traced back to \cite{O54}, \cite{M55}, 
 \cite{T55}, and \cite{P66}, 
but the study of bilinear and trilinear algorithms for MM
 in the papers 
 \cite{W67},
\cite{P72}, \cite{BD73}, \cite{HM73}, \cite{S73}, \cite{P74},
\cite{P76}, \cite{P78}, and \cite{P84}
as well as APA algorithms in the papers 
\cite{BCLR79} and \cite{B80}
have been important steps in this field.
The duality technique for generating 
new efficient bilinear algorithms,
with  applications shown in \cite{W80},
exemplifies other valuable by-products of the MM study.
Besides, as we said earlier, the paper
\cite{P72}
was pioneering in the study of tensor decompositions.
%In particular the border rank of tensors and APA algorithms are
% among the fundamental concepts in that area.
We hope to see new impacts of the MM research 
on the theory and practice of computing in the future.
In the next 
%and concluding 
section we extend the study in \cite{B80}
to some apparently unrelated algorithms for inner products and summation, 
having further link to tensor decomposition.

% - - - - - - - - - - - - - - - - - - - - - - - - - - - - - - - - - - - - -

\section{ APA Approach and
%ractical 
a Summation Algorithm
%and Multilinear 
%Algebra
% under the PRAM Arithmetic Model of Computing
}\label{sbsg}

% - - - - - - - - - - - - - - - - - - - - - - - - - - - - - - - - - - - - -

Recall the APA algorithm of Section \ref{sapa},
assume that the entries $a_{ij}$, $b_{jh}$, $u_{jh}$, and $v_{hi}$
are integers in the range $[0,2^d)$, and choose 
$\lambda=2^d$. Then the 
product $(a_{ij}+\lambda b_{jh})(u_{jh}+\lambda v_{hi})$
would fit the length $L$ of the computer word if $L\ge 4d$.
Moreover if the ratio $L/d$ is large enough, we can perform the
APA computations of Section \ref{sapa} within the
precision $L$. 
\cite[Section 40]{P84b} exploits such observations further 
to devise efficient algorithms for multiplication of
vectors and matrices filled with bounded integers.
Next we recall that technique and  in Example \ref{exsum}
show its surprising extension.
 
%.......................................................

%The paper \cite{B80}
% treats an APA algorithm as a 
%bilinear decompositions whose coefficient are
%polynomial in parameter $\lambda$
%and applies interpolation to recover the desired coefficient
%of this polynomial. Next we follow \cite[Section 40]{P84a} to
% link this technique to some basic
%computations with integers and then apply it 
%to yield a simple but novel algorithm for  fast summation of integers.

Suppose that the coefficient vector 
of a polynomial $v(\lambda)=\sum_{i=0}^{n-1} v_i\lambda^{i}$
is filled with integers from the  semi-open segment $[0,2^d)$
of the real axis for a positive integer $d$.
Represent this vector
by the $2^d$-ary integer
$v(2^d)=\sum_{i=0}^{n-1} v_i2^{di}$. 
Generally the interpolation 
to a polynomial of degree $n-1$ requires its evaluation at $n$ knots,
but in the above special case we  only need the evaluation at the single knot $2^d$.
Now suppose that  all coefficients $v_i$ are integers from the semi-open
segment 
$[q,r)$ for any pair of integers $q$ and $r$, $q<r$.
Then we can apply the above recipe to compute the shifted vector 
${\bf u}=(u_i)_{i=0}^{n-1}=(v_i-q)_{i=0}^{n-1}$, having all its components 
in the  semi-open segment $[0,s)$
for $s=r-q$. Finally we would recover the vector ${\bf v}$ from ${\bf u}$. 
By following \cite{P80a} and \cite{BP86},
we call this technique 
{\em binary segmentation}. Its history
can be traced back to  \cite{FP74},
and one can even view it as 
an application of the Kronecker map,
although
having specific computational flavor. 

Next we follow \cite[Example 40.3, pages 196--197]{P84b} to compute
the inner product of two integer vectors,
then 
extend the algorithm to
summation, and finally 
list  various other applications of
binary segmentation.

\begin{example}\label{exinner}  ({\rm The inner product of two integer vectors}, cf.\ \cite[Example 40.3]{P84b}.)
Assume two nonnegative integers $g$ and $h$ and two vectors ${\bf u}=(u_i)_{i=0}^{n-1}$
and  ${\bf v}=(v_i)_{i=0}^{n-1}$ with nonnegative integer coordinates
 in two semi-open segments, namely, $[0,2^g)$ for the 
coordinates of ${\bf u}$
and $[0,2^h)$ for the coordinates of ${\bf v}$.
%(We can represent the $2n$ input integers by using $(g+h)n$ bits.)
The straightforward algorithm 
for the inner product ${\bf u}^T{\bf v}=\sum_{i=0}^{n-1} u_iv_i$
 at first computes the $n$ products
$u_iv_i$ for $i=0,1,\dots,n-1$
and then sums them. This involves $n$ multiplications and $n-1$ additions.
Instead, however, we can just
 multiply a pair of bounded nonnegative integers, 
%in the semi-open segment $[0,2^{nk})$
apply binary segmentation to the product, and output the 
desired inner product. 
Namely, introduce the two polynomials $u(x)=\sum_{i=0}^{n-1} u_ix^i$
and  $v(x)=\sum_{i=0}^{n-1} v_ix^{n-1-i}$. Their 
product is the polynomial $q(x)=u(x)v(x)=\sum_{i=0}^{2n-2} q_ix^i$ 
%of degree $2n-2$
with integer coefficients in the segment $[0,2^k)$
for $k=g+h+\lceil \log_2 n\rceil$.
The coefficient
$q_{n-1}=\sum_{i=0}^{n-1} u_iv_i$
is precisely the inner product ${\bf u}^T{\bf v}$.
Represent the polynomials $u(x)$ and $v(x)$ by their 
integer values $u(2^k)$ and $v(2^k)$
at the point $2^k$. Clearly, they lie in the semi-open segments 
$r_u=[0,2^{nk+g})$ and $r_v=[0,2^{nk+h})$, 
respectively.
Now compute the integer $q(2^k)=u(2^k)v(2^k)$,
lying in the segment  
$[0,2^{2nk+g+h})$, and 
recover the coefficient $q_{n-1}={\bf u}^T{\bf v}$
by applying binary  segmentation.
\end{example}

\begin{remark}\label{reprod}
We only seek the  coefficient $q_{n-1}$ of the median term 
$q_{n-1}x^{n-1}$
of the polynomial
$u(x)v(x)$. This term lies in  the segment
$[2^{(n-1)k},2^{(n-1)k+g+h})$, and 
the next challenge
is to 
optimize 
its computation. 
Is such a more narrow task substantially simpler 
than the multiplication of two integers lying in the segments
$r_u$ and $r_v$?
\end{remark}

\begin{example}\label{exsum} ({\rm Summation of bounded integers.})
For ${\bf u}=(1)_{i=0}^{n-1}$, $g=0$, and $k=h+\lceil \log_2 n\rceil$ or 
${\bf v}=(1)_{i=0}^{n-1}$, $h=0$, and $k=g+\lceil \log_2 n\rceil$,
the algorithm of Example \ref{exinner} outputs the sum of $n$ 
integers. 
%In this case the  input and  precision are $hn$ and
% $(h+\lceil \log_2 n\rceil)n$, respectively.
\end{example}

\begin{remark}\label{relax}  
In the same way as for polynomial interpolation
in the beginning of this section, we can relax the 
assumption of Examples \ref{exinner} and \ref{exsum}
that the input integers are nonnegative.
Moreover, the summation of integers can be extended to
the fundamental problem of the 
summation of binary numbers truncated to a fixed precision.
% and
% \cite{S82} has already elaborated upon such extensions of binary segmentation.
\end{remark}

%\begin{example}\label{exprod} ({\rm Scaling of an integer vector by an integer.})
%Likewise, we can apply  binary segmentation to multiply
%a vector ${\bf v}=(v_i)_{i=0}^{n-1}$ by a scalar $u$
%where all $v_i$ and $u$ are integers, $u$ segments  
%from $1-2^g$ to $2^g-1$ 
%and  $v_i$ range from $1-2^h$ to $2^h-1$ for all $i$, for two nonnegative integers $g$ and $h$.
%Now define the polynomial  $v(x)=\sum_{i=0}^{n-1} v_ix^{n-1-i}$,
%map the vector ${\bf v}$ into the integer $v(2^k)$
%for $k=g+h$,
%multiply this integer by the integer $u$,
%and recover the products $uv_0,\dots,uv_{n-1}$ from the
% representation of the integer  
%$uv(2^k)=(uv_{n-1}~|~ uv_{n-1}~|~\dots~|~uv_1~|~uv_0)_{2^k}$
%with the basis $2^k$. In this case binary segmentation
%increases the overall input precision from $g+hn$ to 
%$g+(g+h)n$.
%\end{example}

In Examples \ref{exinner} and \ref{exsum},
 multiplication of
%with bounded integers to a single multiplication of
 two long integers followed by binary segmentation replaces
 either $2n-1$ or $n$ arithmetic operations,
respectively.
This increases the
 Boolean (bit-wise operation) cost 
by a factor depending on the Boolean cost 
of computing the
 product of 2 integers or, in view of
Remark \ref{reprod}, of computing the  
median segment in the binary representation
of the product.
The increase is minor if we multiply integers in 
nearly linear Boolean time (see the supporting algorithms for such multiplication in
  \cite{SS71}, \cite[Section 7.5]{AHU74}, \cite{F09}),
but grows if we multiply integers
by applying the straightforward algorithm, 
which uses quadratic Boolean time.
Nonetheless, in both cases
 one could still benefit from using Example  \ref{exsum}
if
the necessary bits of the output integer fits the computer word
(i.e. the bits of the middle coefficient are not part of the overflow of the product),
%e.g., the IEEE standard double precision of computing or the integer
%representation Int64,
as long as the representation of the vector as an integer
requires no additional cost.
% (one can formally estimate the benefit by assuming the RAM arithmetic model
% for a fixed word length 
% \cite{AHU74}).
If
  the output
integer  does not fit the 
word length, 
we can apply the same algorithms to the subproblems of smaller sizes,
e.g., we can apply the algorithms of Examples \ref{exinner}
and  \ref{exsum} to
compute the inner products of some subvectors
 or partial sums of integers, respectively.
%Moreover, even if the present day computers are unfavorable to
%our applications of binary segmentation,
%this can change for the computers of the future. 

Other applications of binary segmentation include
 polynomial multiplication (that is, the computation of  the 
convolution of vectors)   
 \cite{FP74}, \cite{S82},  some basic linear algebra
 computations \cite[Examples 40.1--40.3]{P84b}, 
  polynomial division  \cite{BP86}, \cite{S82},
 computing polynomial GCD   \cite{CGG84},
and  discrete Fourier transform 
 \cite{S82}. Binary segmentation can be potentially efficient in  
computations with Boolean vectors and matrices.
E.g., recall that Boolean MM is reduced to MM whose input and output entries 
are some bounded nonnegative integers (see  \cite[Proof of Theorem 6.9]{AHU74}). 
Quantized  
tensor decompositions is another promising application area (cf.\   
 \cite{T03}, \cite{O09},
\cite{O10}, \cite{K11}, \cite{OT11}, \cite{GKT13}).

{\bf Acknowlegements.} 
My work has been supported by 
 NSF Grant CCF--1116736 and
PSC CUNY Award 67699-00 45.
Furthermore I am grateful to A. Bostan, I.V. Oseledets and E.E. Tyrtyshnikov
for their valuable pointers to some important recent works on MM and the bibliography 
 on  quantized tensor decompositions, respectively, to
Franklin Lee, Tayfun Pay, and Liang Zhao for their helpful comments
on potential implementation of 
the algorithms of Examples \ref{exinner} and \ref{exsum},
and excellent assistance with creating
Figures \ref{fig1} and \ref{fig2}, and to Ivo Hedtke for his very
helpful comments to my original draft.

%------------------------------------------------------------------------------

%------------------------------------------------------------------------------
%------------------------------------------------------------------------------

\end{document}